\documentstyle[epsfig,longtable]{aipproc}
\newcommand{\be}{\begin{equation}}
\newcommand{\ee}{\end{equation}}
\begin{document}

\title{Disk Instabilities and Cooling Fronts}
 
\author{Ethan T. Vishniac}
\address{Dept. of Astronomy, University of Texas, Austin, TX 78712}

%\lefthead{LEFT head}
%\righthead{RIGHT head}
\maketitle

\begin{abstract}
Accretion disk outbursts, and their subsequent decline, offer a
unique opportunity to constrain the physics of angular momentum
transport in hot accretion disks.  Recent work has centered
on the claim by Cannizzo et al. that the exponential decay of 
luminosity following an outburst in black hole accretion disk
systems is only consistent with a particular
form for the dimensionless viscosity, $\alpha=35(c_s/r\Omega)^{3/2}$.
This result can be understood in terms of a simple model
of the evolution of cooling fronts in accretion disks. 
In particular, the cooling front speed during decline is 
$\sim \alpha_F c_{s,F}(c_{s,F}/r\Omega)^{n}$, where
$F$ denotes the position of the cooling front, and
the exact value of $n$ depends on the hot state opacity,
(although generally $n\approx 1/2$).  Setting this speed
proportional to $r$ constrains the functional form of $\alpha$
in the hot phase of the disk, which sets it apart from previous
arguments based on the relative durations of outburst and
quiescence.  However, it remains uncertain how well we know the 
exponent $n$.  In addition, more work is needed to clarify the
role of irradiation in these systems and its effect on the
cooling front evolution.
\end{abstract}

\section*{Introduction}

The most popular model for soft X-ray transients and dwarf novae
is that both are due to thermal instabilities in the accretion disks
around collapsed objects.  The specific mechanism thought
to drive the thermal instability is the change in opacity associated
with the ionization of hydrogen (for a recent review see
Cannizzo 1993\cite{c93}, or Osaki 1996\cite{o96}).  In these systems
the accretion disks can be modeled as geometrically thin, $H\ll r$, 
and with vertical gravity supplied by the central object.  As a
result the sound speed to orbital velocity ratio, $c_s/(r\Omega)$
scales with the thickness to radius ratio, $H/r$, and the
orbital frequency $\Omega(r)$ is proportional to $r^{-3/2}$.

The outburst cycle in these systems consists of alternating
periods of quiescence and outburst.  During
quiescence the disk luminosity is low and gas is
accreted from the companion without being transferred through the disk.
In this state the disk is mostly neutral and cold.  Eventually the
accumulation of material at large radii leads to the runaway
ionization of hydrogen and the disk material jumps to a hot
state.  The mass transfer rate increases dramatically and a heating
wave propagates inward at the thermal speed, $\sim \alpha c_s$,
where $\alpha$ is the dimensionless disk viscosity and $c_s$ is
the midplane sound speed.  In a short time the entire disk
is ionized and hot, and remains in that state until the steady
depletion of material makes it impossible to sustain the outburst.
At this point the disk starts to fall back into the cold state,
usually starting at some large radius, and a cooling wave sweeps
inward while the luminosity of the disk decreases.  The duration
of the outburst and quiescent phases is consistent with an
outburst value of $\alpha$ close to $0.1$ and a cold state value
several times smaller.  

One of the more striking features of these systems is the
extent to which the luminosity decline can be modeled by
an exponential decay law, in extreme cases stretching of three orders
of magnitude in X-ray luminosity  (Chen et al. 1997\cite{csl97})
Mineshige et al. (1993)\cite{myi93} noted that this implies an
exponential decay of the disk hot phase, and that this can be achieved
in the context of the disk instability model by requiring the
cooling wave velocity to scale with the radius of the hot phase.
More recently Cannizzo, Chen \& Livio (1995)\cite{ccl95}  have explored this argument
through a series of high resolution simulations, and found that
this scaling can be obtained only for a restricted class of
viscosity laws, and that models with a constant $\alpha$
can be strongly excluded.  Here we will review recent progress
on the connection between cooling waves and models for angular
momentum transport in accretion disks.

\section*{The Cooling Front Speed}

If the hot, inner parts of the accretion disk are evolving exponentially,
then the simplest way to drive this evolution is to have the boundary
of the hot phase of the disk evolve exponentially, that is
\be
{V_F\over r_F}=\hbox{constant},
\ee
where the subscript $F$ denotes the boundary of the hot phase, where
rapid cooling sets in due to the thermal instability, and
$V_F\equiv\dot r_F$.  (Note that this may not be the {\it only}
way to drive an exponential decay in $L_x$.)  In any case, the
requirement that $V_F$ be proportional to $r_F$ immediately forces
a connection between the local physics of the cooling front, and the
global evolution of the disk.

The earliest work on cooling front evolution (Mineshige 1987\cite{m87}, 
Cannizzo, Shafter, and Wheeler 1988\cite{csw88}) led to the conclusion
that
\be
V_F\sim \alpha_F c_{s,F} \left({\Delta r\over r}\right).
\label{eq:vfold}
\ee
In other words, that the cooling front speed is the local cooling rate,
$\alpha\Omega$, times the local disk height modified by a universal 
spreading factor, thought to be $\sim 0.1$.  This result reflects the
view that the cooling front should be thought of as a purely local
phenomena.

However, recently CCL\cite{ccl95} 
showed instead that the cooling front speed can be approximated as
\be
V_F\sim \alpha_F c_{s,F}\left({c_{s,F}\over r_F\Omega_F}\right)^{1/2}.
\label{eq:vfnew}
\ee
The significance of this result follows from its implications for
$\alpha$.  Since the midplane disk temperature at the cooling front
is a very weak function of radius, equation (\ref{eq:vfold}) implies
that $\alpha_F\propto r_F$, whereas equation (\ref{eq:vfnew}) implies
$\alpha_F\propto r_F^{3/4}$.  Cannizzo et al. suggested adopting
the rule
\be
\alpha=\alpha_0\left({c_s\over r\Omega}\right)^{3/2},
\label{eq:alpha}
\ee
following the form proposed by Meyer and Meyer-Hofmeister (1983)\cite{mmh83}.  The
time scale for the luminosity decay is then 
\be
\tau_F\approx 0.4{GM_{BH}\over\alpha_0 c_{s,F}^3}.
\label{eq:decay}
\ee
Matching the observations they found
\be
\alpha_0\approx 35{M_{BH}\over 7 M_{\sun}}.
\label{eq:alpha0}
\ee
They showed that this prescription leads to reasonable agreement with observations
of X-ray novae, and were unable to find any other simple prescription that did as well.
An added advantage of this model is apparent from equation (\ref{eq:decay}).
One immediately obtains dramatically shorter time scales for dwarf novae,
in agreement with observations.  In fact, the same prescription can be used to
model the outburst cycle of SS Cygni (Cannizzo 1996\cite{c96}).

These results raise several important questions.  First, why is there a 
factor of $(c_s/r\Omega)^{1/2}$ (or $(H/r)^{1/2}$) in equation (\ref{eq:vfnew})?
There are only two obvious velocities in the problem, the accretion velocity,
$\alpha c_s^2/(r\Omega)$, and the thermal speed, $\alpha c_s$.  Is this
some sort of geometric average or is it totally unrelated?  Second, is
equation (\ref{eq:alpha}) another way of describing the drop in $alpha$
from the hot state to the cold state, or is it a measurement of $\alpha$
in the hot state alone?  Third, if this is the correct form for $\alpha$,
why is the  exponent  $3/2$?  Finally, the coefficient $\alpha_0$ is
presumably a constant of order unity.  However, $35$ is a peculiarly
large value for the number one.  Where does this number come from?   

To anticipate the rest of this presentation,  we now have an analytic theory
for cooling waves in a locally heated disk (Vishniac
and Wheeler 1996\cite{vw96}, Vishniac 1997\cite{v97}) and the first two questions can 
be answered.  The third and fourth questions remain confusing.

The analytic theory of cooling waves starts from the three equations which
describe a vertically averaged disk model:
\be
\partial_t\Sigma ={-1\over r}\partial_r(r\Sigma V_r),
\ee
\be
V_r={2\over\Sigma\Omega r^2}\partial_r(r^3\alpha\Sigma c_s^2\partial_r\ln\Omega),
\label{eq:torque}
\ee
and an equation for thermal structure, which we take to be
\be
T=B\Sigma^a\alpha^b\Omega^{{2\over3}c},
\label{eq:thermal}
\ee
following CCL, where $a=3/7$, $b=1/7$, and $c=3/7$. We need to add 
the condition that below
some critical temperature, which is a very weak function of radius, cooling
becomes rapid.  At this radius the radial thermal
gradient becomes very large.  This in turn implies
that a large fraction of the angular momentum flux from smaller radii is deposited, 
following equation (\ref{eq:torque}) within a radial scale length, $r_T$.  The
resulting gas velocity is 
\be
V_r\sim{\alpha_F c_{s,F}^2\over r_T\Omega_F}.
\ee
Since the gas cools at a rate $\sim \alpha_F\Omega_F$, we can estimate $r_T$ as
\be
r_T\sim {V_r\over\alpha_F\Omega_F}.
\ee
We conclude that $r_T$ is some constant times the disk height and $V_r$ is
directed outward.

This rapid outward motion at the cooling front implies that the cooling front
is preceded by a rarefaction wave.  This suggests that we can understand
the progress of the cooling wave by considering the dynamics of the rarefaction
wave.  If we label a point just ahead of the rarefaction wave with a `p',
then conservation of mass implies that
\be
\Sigma_p V_F\approx \Sigma_F\alpha_F c_{s,F},
\ee
or
\be
V_F\approx \alpha_F c_{s,F} {\Sigma_F\over\Sigma_p},
\label{eq:vf1}
\ee
assuming that the cooling front speed is more rapid than the radial motion
of the gas at point `p'.  The column density $\Sigma_F$ is related to
the critical cooling temperature through equation (\ref{eq:thermal}).  
The column density $\Sigma_p$ needs to be constrained separately to
estimate $V_F$.  We can do this by setting $V_F$ equal to the
accretion velocity at the precursor point.  In other words,
\be
V_F\approx\alpha_p{c_{s,p}^2\over r\Omega}\sim\alpha_F c_{s,F}{\Sigma_F\over\Sigma_p}.
\label{eq:vf2}
\ee
Our rationale for this choice is that no other choice is self-consistent.  If
$V_F$ is much faster than this, then the accretion disk does not evolve significantly
as the cooling wave plows through it.  This decouples the cooling wave behavior
from the evolution of $\dot M$, but it also implies that $\Sigma_p$ climbs sharply
as the cooling wave moves in.  If we remember that $T_F$ is nearly constant,
then equation (\ref{eq:thermal}) implies that $\Sigma_F$ actually drops as
at smaller radii.  This implies a steady decrease in $\Sigma_F/\Sigma_p$, and
a consequent drop in $V_F$ (following equation (\ref{eq:vf1}).  On the other
hand, if $V_F$ is much smaller than the accretion velocity ahead of the rarefaction
wave, then the disk evolves faster than the cooling wave can move.  This is
paradoxical since the draining of the inner disk automatically drops the disk 
into the cold state.

We can use equations (\ref{eq:thermal}) and  (\ref{eq:vf2}) to solve for $V_F$
for any assumed form for $\alpha$.
If 
\be
\alpha=\alpha_0\left({c_s\over r\Omega}\right)^n,
\ee
then
\be
V_F \approx \alpha_F c_{s,F}\left({c_{s,F}\over r\Omega}\right)^q,
\label{eq:vfinal}
\ee
where
\be
q^{-1}=1+\left(1+{n\over 2}\right){a\over 1-(n/2)b}.
\ee
For a Kramers opacity law, the usual choice for the disk hot state,  and $n=3/2$ 
this gives $q=25/46$.  In general, it is very difficult to get a value of
$q$ which is very different from 1/2 for any of the usual opacity laws.

What have we learned from this scaling law argument?  First,
this is an observational test for the functional form of $\alpha$ 
{\it in the hot state}, in 
spite of the fact that a cooling transition is involved.  The dynamics of the 
wave are controlled by the rarefaction wave in the hot material, and do not
depend on the structure of the rapid cooling region.  There is a possible
loophole here, which we will return to later.  The cold material will pile
up after it has finished cooling, with a column density not much below
$\Sigma_p$.  We have assumed that this is not more than the maximum column
density on the lower branch of the `S' curve.
If this condition is not satisfied, then gas is not actually ejected
at a constant fraction of the thermal speed across the cooling front and 
we have, instead,
\be
V_F\sim\alpha_F{c_{s,F}^2\over r\Omega}\left({\Sigma_{hot,minimum}\over\Sigma_{cold,maximum}}\right)
^{1-1/q}.
\ee
As the `S' curve central branch becomes vertical this gives a cooling front
velocity which is close to the accretion speed.

Second, as a test of the model we can ask how it behaves if
the upper branch of the `S' curve is almost horizontal, i.e. if the
temperature is insensitive to the column density.  In this limit, $a=0$ 
and consequently $q=1$. As expected, the cooling front moves at the accretion 
velocity.   In the opposite limit, when the hot state is just barely
thermally stable, $a\rightarrow\infty$ and $q\rightarrow 0$.  That is,
the cooling front moves at close to the thermal speed.  We can understand
the intermediate behavior found by CCL as a reflection of the fact that
realistic disks have a hot state which lies between these two extreme
cases.

Third, we note that once we allow for the variation of the minimum hot
state temperature with radius, a perfect exponential decline actually follows 
for $n=1.63$.  This has been confirmed by subsequent numerical tests (Cannizzo 1997
\cite{c97}).  
This does not imply that this is the correct value for the exponent $n$, since
the observations aren't necessarily perfectly exponential.  Instead, this
should be regarded as test of the ability of the scaling law to predict the
results of numerical simulations.

We can improve on this scaling law by constructing a similarity solution
for the disk gas using the equations of continuity, torque, and thermal structure
(Vishniac 1997\cite{v97}).  The result is only an approximate fit to the actual
solution, since the problem itself has a built-in scale.  The appropriate
boundary condition for $\dot M$ at the front is
\be
\dot M_F=2\pi r_F\Sigma_F \left(\alpha_F c_{s,F} \Delta\right),
\ee
where $\Delta$ is a constant relating the outflow speed at the front to the thermal
speed at the front.  The analytic theory does not constrain the value
of $\Delta$ so we need to take it from the simulations of Cannizzo et al.. 
(This gives $\Delta\approx 1/6$).
The second boundary condition looks absurd. It is 
\be
\Sigma_F=0.
\ee
In practice this means only that the density contrast through the rarefaction wave
is large.

From this we derive a similarity solution with
\be
{\partial_t \dot M_{inner}\over \partial_t \ln r_F}=2.37,
\ee
and
\be
V_F=0.94 \alpha_F c_{s,F} \left({c_{s,F}\over r\Omega}\right)^q(6\Delta)^{1/2}.
\ee
Both of which are consistent with the simulation results.  At any one time
more than half the hot phase of the disk is actually moving outward, although
the radial speed is close to the usual accretion velocity except in a thin
annulus close to the cooling front.  Comparing the similarity solution
to the numerical work we find that the two differ by about
5-10\% which gives us confidence that we understand both of them.

\section*{Implications for a Theory of Angular Momentum Transport}

The model presented here for the evolution of the cooling wave
is based on assuming that the disk is heated through the local
dissipation of orbital energy and that $\alpha$ can be described
in terms of local variables.  Assuming for the moment that this
is basically correct, what are the implications for a fundamental
theory of $\alpha$? Knowing that a particular functional form of $\alpha$
fits the available observational evidence falls well short of
a fundamental theory for $\alpha$.  On the other hand, we can use
our results to identify particularly promising models, if any exist.
Since we are concerned here with angular momentum transport in the hot, ionized phase
we can restrict our attention to models which are most efficient in this phase.

All currently popular models for this regime make use of the Velikhov-Chandrasekhar
instability (Velikhov 1959\cite{v59}, Chandrasekhar 1961\cite{c61}), 
first identified by Balbus and Hawley (1991)\cite{bh91} as the most,
and perhaps only, promising mechanism for angular momentum transport in 
conducting accretion disks.  This is an instability of a magnetic field embedded
in a strongly shearing flow.  The linear instability has a typical growth rate $\sim\Omega$ 
and a typical length scale $\sim V_A/\Omega$ with a resultant $\alpha\propto V_A^2/c_s^2$.  
Numerical simulations (cf. the review by Gammie 1997\cite{g97} 
and references therein) confirm
the existence of the instability and its saturation in a turbulent state.  Interestingly
enough, they also indicate that vertical structure is largely irrelevant
for the dynamo process, which allows us to exclude magnetic buoyancy
as a driving force for the dynamo.  Differences
among the models based on these results are due entirely to different proposals for
the underlying dynamo mechanism.

First, the turbulence might support an inverse cascade, in which a large scale
field is generated spontaneously from the underlying turbulence without any
appeal to the kind of symmetry breaking used in mean-field dynamo theory
(Balbus and Hawley 1991\cite{bh91}, see also Gammie 1997\cite{g97} 
and references therein).  
Since this is a purely local process, it produces an $\alpha$ which is 
some constant of order unity, with a relaxation time which is some constant
times the eddy turn over time ($\sim \Omega^{-1}$).  The available simulations 
seem consistent with this model over a limited range of box sizes, but the
resultant $\alpha$ is a function of grid resolution (e.g. Brandenburg et al.
1996\cite{bnst96}).  Periodic shearing
box sizes have a length which varies between $1$ and $4$ times $2\pi$ times the
box height, which is probably a measure of the sensitivity to disk geometry
over the range $1<(r/H)<4$ in a thin disk.  
However, this model is inconsistent with the cooling wave results.
If the latter are correct, then the model needs to be modified to allow some
non-local effects.

Second, there is the internal wave-driven dynamo (Vishniac \& Diamond 1992\cite{vd92},
and references therein) in which tidally generated waves drive a dynamo with 
a growth rate $\sim (H/r)^{n}$,
where $n$ is a bit less than $1.5$ in a stationary disk.  
Balancing dissipation with growth gives $\alpha\sim \Gamma/\Omega$.
This is the right scaling, but this dynamo model is inherently nonlocal in
a bad sense.  The dependence on internal waves propagating through the cool region should 
make this dynamo {\it very} weak just inside the cooling front, wiping out the `S' curve effect 
and completely changing the evolution of the cooling front.  Although no detailed
calculation of these effects exist, it seems likely that this would lead to a broad cooling
front, propagating at the viscous speed, with a dependence on $(H/r)$ which is
too steep.

Finally, there has been an attempt to construct a mean field dynamo theory with
no symmetry breaking, the incoherent mean field dynamo (Vishniac and Brandenburg 1997
\cite{vb97}).
In this model the helicity necessary for generating  large scale radial magnetic
field emerges from the random fluctuations provided by the small scale instabilities
of the magnetic field.  It can be shown that this process will drive the exponential
growth of large scale, axisymmetric magnetic field domains in the presence of
strong shearing.  However, this process becomes less efficient as the number of
eddies per annulus increases, i.e. this dynamo runs more slowly as $H/r$ decreases
(or, in simulations, as the height to length ratio of the periodic shearing box
decreases).  In a sense, this is an attempt to implement a geometry dependent
inverse cascade in the context of the first model.  Unfortunately, while this
model seems qualitatively promising, it predicts
\be
\alpha\sim \left({V_A\over c_s}\right)^2\propto \left({H\over r}\right)^2,
\label{eq:scaling}
\ee
which is too steep.  Moreover, the numerical simulations do not show the expected
sensitivity to geometry.  This may be because the large value of $\alpha_0$
restricts the asymptotic regime, where the scaling law given in equation (\ref{eq:scaling})
is valid, to boxes considerably longer and thinner than the range explored to date.
In any case, in order to reconcile this model to the cooling wave calculations, we
require dwarf novae and X-ray novae disks to be sufficiently thick that small
corrections to the dynamo growth soften the saturated value of $\alpha$.  More
detailed modeling of this dynamo process will be necessary to judge whether or not
this is a realistic hope.

We conclude that no current model is fully consistent with the cooling wave results,
although there is some chance that a more realistic version of the incoherent 
dynamo model will prove adequate.  

\section*{Future Prospects}

What can we expect by way of progress in this area?  Our ability to set limits
on the allowed forms of $\alpha$ is limited by the range of luminosities we
can observe and the presence of non-exponential features in accretion disk
light curves, e.g. reflares and secondary maxima.  As long as these features
remain poorly understood it will be difficult to pin down the required form
of $\alpha$.  However, even within the limits imposed by our current state of
understanding the computational models could be made to yield significantly
more information.  For example, we have no precise idea of the appropriate
error bars are on the parameters of the proposed functional form of $\alpha$.
CCL originally estimated that the value of $n$ allowed by observational
constraints was probably within $15\%$ of the `exponential' value of  $1.5$, but the 
value of $n$ which actually yields a perfect exponential decay is $10\%$ higher,
and the available grid of models has not yet nailed down the allowed range
of $n$.  

A more serious danger is the possibility is that the cooling wave model used here 
neglects some important physical effects.  The relaxation of the magnetic
field as the cooling wave approaches is difficult to calculate, since the
actual dynamo mechanism is, as noted above, a controversial topic.  However,
we expect the typical rate for this relaxation to be comparable to
the dynamo growth rate, which should be $\sim\alpha\Omega$.  This is also
the thermal relaxation rate in an optically thick disk.  The disk structure
will evolve at slower rates until the onset of rapid cooling, which is
too late to affect the cooling wave arguments.  Moreover, even if
we assume that the magnetic field is frozen into the plasma from the very
beginning of the rarefaction wave we can show that it has no perceptible
effect on our results.

Another, and potentially more serious problem is that this treatment assumes
that the disk temperature is determined by local dissipation, that is, it
ignores irradiation.  Attempts to account for the optical light from BH
systems suggest that irradiation may be as significant as local energy
dissipation (Cannizzo 1997\cite{c97}), although it probably does not dominate by
a large factor.  If we restrict our attention to dwarf novae systems, where
irradiation is less important, then we find similar, but less
precise, constraints on $\alpha$ (Canizzo 1996\cite{c96}).  In any case, the effect of
moderate irradiation on the cooling wave structure needs to be explored.

Are there acceptable alternatives to the cooling wave model for the
these systems, perhaps ones that not involve a varying $\alpha$?
At present the answer is no, although a thorough search has
not yet been done.
If we wish to consider alternative explanations for the exponential decay
of X-ray emission from black hole accretion disk systems, we need models that
maintain a constant accretion time scale in the relevant mass reservoir,  
that is
\be
\tau^{-1}_{acc}=\alpha{c_s^2\over r^2\Omega}=\hbox{constant},
\label{eq:tau}
\ee
at the outermost radius which is feeding matter through the hot part
of the disk.  In the cooling wave model this is the precursor point,
and the somewhat odd form for the cooling wave speed given in equation
(\ref{eq:vfinal}) is a consequence of the fact that the precursor
point is not at the cooling front.  A model in which irradiation
maintains the entire disk in a hot state, so that $r$ is the tidal
boundary of the disk, will satisfy equation (\ref{eq:tau})
only if $\alpha\propto T^{-1}$.  A model dominated by irradiation
in which the outer edge of the hot phase is defined by a constant
temperature, and the dynamics of the cooling wave are suppressed,
will work only if $\alpha\propto (H/r)$, which differs from the
cooling wave constraint, but is just as problematic in terms of
the fundamental physics of $\alpha$.  The only obvious way to combine
an exponential decay law with a constant $\alpha$ is to present
a model in which both the temperature and radius at the disk edge
are nearly constant.  This would require a transfer of energy from
the inner disk whose efficiency increases sharply during the decline,
perhaps as a result of coronal scattering through an inner disk 
corona whose size and density increases smoothly as $L_x$ decreases.
It remains to be seen whether or not a physically reasonable model
can be constructed along these lines.

This work has been supported in part by NASA grant NAG5-2773.

\end{document}